\documentclass[12pt]{article} 

\usepackage[utf8]{inputenc} 
\usepackage[english]{babel}
\usepackage{hyperref}
\usepackage{latexsym}
\usepackage{amsmath}
\usepackage{amsfonts}
\usepackage{times}
\usepackage{xcolor}
\usepackage{graphicx}
\usepackage{xspace}
\usepackage{algorithmicx}
\usepackage{natbib}
\usepackage{pdflscape}
\usepackage{kbordermatrix}

\newcommand*{\affaddr}[1]{#1} 
\newcommand*{\affmark}[1][*]{\textsuperscript{#1}}
\newcommand*{\email}[1]{\texttt{#1}}

\title{\LARGE\textbf{On Fractional Approach\\ to Analysis of Linked Networks}}

\author{
Vladimir Batagelj\affmark[1,2,3]  \\
ORCID: 0000-0002-0240-9446\\
\affaddr{\affmark[1]Institute of Mathematics, Physics and Mechanics,\\ Jadranska 19, 1000 Ljubljana, Slovenia}\\
\affaddr{\affmark[2]University of Primorska, Andrej Marušič Institute, 6000 Koper, Slovenia}\\ 
\affaddr{\affmark[3] National Research University Higher School of Economics,\\  Myasnitskaya, 20, 101000 Moscow, Russia.}\\
\email{vladimir.batagelj@fmf.uni-lj.si}\\ phone: +386 1 434 0 111
}

\newcommand{\keyw}[1]{\textcolor{red}{\emph{#1}}}

\newcommand{\WA}{\mathbf{W\!\!A}}

\newcommand{\WK}{\mathbf{W\!K}}

\newcommand{\WJ}{\mathbf{W\!J}}
\newcommand{\AK}{\mathbf{A\!K}}
\newcommand{\Ci}{\mathbf{Ci}}
\newcommand{\Co}{\mathbf{Co}}

\newcommand{\Hm}{\mathbf{H}}

\newcommand{\network}[1]{\mathcal{#1}}
\newcommand{\vertices}[1]{\mathcal{#1}}
\newcommand{\edges}[1]{\mathcal{#1}}

\newcommand{\outdeg}{\mbox{outdeg}}
\newcommand{\indeg}{\mbox{indeg}}

\newcommand{\diag}{\mbox{diag}}

\newcommand{\Mw}{\mathop{\raisebox{-1.5pt}{\mbox{$\Box$\kern-.55em\raisebox{2.5pt}{{\tiny $r$}}\kern2.9pt}}}}
\newcommand{\Mv}{\mathop{\raisebox{-1.5pt}{\mbox{$\Box$\kern-.55em\raisebox{2.5pt}{{\tiny $h$}}\kern2.9pt}}}}

\newcommand{\Remark}[1]{\ifodd\value{page} \normalmarginpar
 \else \reversemarginpar \fi \marginpar{{\footnotesize #1}} }

\newcommand{\clock}{\count254=\time \divide\count254 by 60
 \count255=\count254 \multiply\count255 by -60
 \advance\count255 by \time
 \ifnum\count254<10 0\fi\number\count254\,:\,%
 \ifnum\count255<10 0\fi\number\count255}

\graphicspath{{./}{./pics/}}

\begin{document}

\hypersetup{pdfauthor={V. Batagelj}}
\hypersetup{pdftitle={On Fractional Approach to Analysis of Linked Networks}}

\maketitle

\begin{abstract}
In this paper, we present the outer product decomposition of a product of compatible linked networks. It provides a foundation for the fractional approach in network analysis. We discuss the standard and Newman's normalization of networks. We propose some alternatives for fractional bibliographic coupling measures.
\\[4pt]
\\[4pt]
\\[4pt]

\end{abstract}


\section{Introduction}

The fractional approach was proposed by \cite{fracL}. For example in the analysis of coauthorship the contributions of all coauthors to a work has to add to 1. Usually the contribution is then estimated as 1 divided by the number of coauthors. An alternative rule, Newman's normalization, was given in \cite{newman1} and \cite{newman4} which excludes the selfcollaboration.
Recently several papers \citep{fracBC,fracCB,fracPWE,fracPM,fracLP,fracG} reconsidered the background of the fractional approach. The details are presented and discussed in Subsection~\ref{fa}. In this paper we propose a theoretical framework based on  the outer product decomposition to get the insight into the structure of bibliographic networks obtained with network multiplication.


\section{Linked networks}

Linked or multi-modal networks are collections of networks over at least two sets of nodes (modes)
and consist of some one-mode networks and some two-mode networks linking different modes.
For example: modes are Persons and Organizations. Two one-mode networks describe collaboration
among Persons and among Organizations. The linking two-mode network describes membership of
Persons to different Organizations. 

Linked networks are the basis of the MetaMatrix approach developed by Krackhardt and Carley \citep{linkKC,linkC}. For an example see the Table 3 in \citet[p. 89]{MeMa}.

Another example of linked networks are bibliographic networks.
From special bibliographies (\href{http://www.math.utah.edu/~beebe/}{Bib\TeX})
and bibliographic services
(\href{http://thomsonreuters.com/products_services/science/science_products/a-z/web_of_science/}{Web of Science},
\href{http://www.scopus.com/home.url}{Scopus},
\href{http://sicris.izum.si/default.aspx?lang=eng}{SICRIS},
\href{http://citeseer.ist.psu.edu/}{CiteSeer},
\href{http://www.zentralblatt-math.org/zmath/}{Zentralblatt MATH},
\href{http://scholar.google.com/schhp?hl=en}{Google Scholar},
\href{http://www.informatik.uni-trier.de/~ley/db/}{DBLP Bibliography},
\href{http://www.uspto.gov/}{US patent office},
\href{http://www.imdb.com/interfaces}{IMDb},
and others)
we can construct some two-mode networks on selected topics:
authorship on works $\times$ authors ($\WA$),
keywordship on works $\times$ keywords ($\WK$),
journalship on works $\times$ journals/publishers ($\WJ$),
 and from some data also the classification network on
works $\times$ classification ($\mathbf{WC}$) 
and the one-mode citation network on works $\times$ works ($\Ci$);
where works include papers, reports, books, patents, movies, etc.
Besides this we get also the partition of works by the publication year, and the vector of number of pages \citep{wos,wos2pajek}.

An important tool in analysis of linked networks is the use of derived networks obtained by network multiplication.


\section{Network multiplication}

Given a pair of \keyw{compatible} two-mode networks
$\network{N}_A = (\vertices{I},\vertices{K},\edges{A}_A,w_A)$ and
$\network{N}_B = (\vertices{K},\vertices{J},\edges{A}_B,w_B)$ with
corresponding matrices $\mathbf{A}_{\vertices{I} \times \vertices{K}}$
and $\mathbf{B}_{\vertices{K} \times \vertices{J}}$
we call a \keyw{product of networks} $\network{N}_A$ and $\network{N}_B$ a network
$\network{N}_C = (\vertices{I},\vertices{J},\edges{A}_C,w_C)$,
where $\edges{A}_C = \{ (i,j): i \in \vertices{I}, j \in \vertices{J}, c_{i,j} \ne 0 \}$
and $w_C(i,j) = c_{i,j}$ for $(i,j) \in \edges{A}_C$. The product matrix
 $\mathbf{C} = [ c_{i,j} ]_{\vertices{I} \times \vertices{J}} = \mathbf{A} \cdot \mathbf{B}$
is defined in the standard way
\[ c_{i,j} = \sum_{k \in \vertices{K}} a_{i,k} \cdot b_{k,j} \]
In the case when $\vertices{I} = \vertices{K} = \vertices{J}$ we are dealing with ordinary one-mode
networks (with square matrices).

In the following we will often identify networks by their matrices.

In the paper \citet{fracBC} it is shown that $c_{i,j}$ is equal to the value of all two step paths  from $i \in \vertices{I}$
to $j \in \vertices{J}$ passing through $\vertices{K}$. In a special case,
if all weights in networks $\network{N}_A$ and $\network{N}_B$ are equal to 1 the
value of $c_{i,j}$ counts the number of ways we can go from $i \in \vertices{I}$
to $j \in \vertices{J}$ passing through $\vertices{K}$: $c_{i,j} = | N_A(i) \cap N^-_B(j)| $;
where $N_A(i)$ is the set of nodes in $\vertices{K}$ linked by arcs from node $i$ in the network $\network{N}_A $, and $N^-_B(j)$ is the set  of nodes in $\vertices{K}$ linked by arcs to node $j$ in the network $\network{N}_B $.

The standard matrix multiplication has the complexity
$O(|\vertices{I}|\cdot |\vertices{K}|\cdot |\vertices{J}|)$ -- it is
too slow to be used for large networks.
For sparse large networks we can multiply much faster considering only
nonzero elements.
\begin{tabbing}
xxx\=xxx\=xxx\=xxx\=xxx\kill
\>\textbf{for} $k$ \textbf{in} $\vertices{K}$ \textbf{do} \\
\>\> \textbf{for} $(i,j)$ \textbf{in} $N^-_A(k) \times N_B(k)$ \textbf{do}  \\
\>\>\> \textbf{if} $\exists c_{i,j}$ \textbf{then}
         $c_{i,j} := c_{i,j} + a_{i,k} \cdot b_{k,j}$ \\
\>\>\> \textbf{else} new $c_{i,j} := a_{i,k} \cdot b_{k,j}$
\end{tabbing}

In general the multiplication of large sparse networks is a
'dangerous' operation since the result can
'explode' -- it is not sparse.
If for the sparse networks $\network{N}_A$ and
$\network{N}_B$ there are in $\vertices{K}$ only few nodes  with large
degree and no one among them with large degree in both networks
then also the resulting product network $\network{N}_C$ is sparse.

From the network multiplication algorithm we see that each intermediate node
$k \in \vertices{K}$ adds to a product network a complete two-mode subgraph
$K_{N^-_A(k),N_B(k)}$ (or, in the case $\mathbf{B} = \mathbf{A}^T$, where $\mathbf{A}^T$ is the transposition of $\mathbf{A}$, a complete
subgraph $K_{N(k)}$). If both degrees $\deg_A(k)=|N^-_A(k)|$
and $\deg_B(k)=|N_B(k)|$ are large then
already the computation of this complete subgraph has a quadratic (time and space)
complexity -- the result 'explodes'.
For details see the paper \citet{fracBC}.

\section{Outer product decomposition}

\begin{figure}
\begin{center}
\includegraphics[width=65mm,viewport=170 80 625 560,clip=]{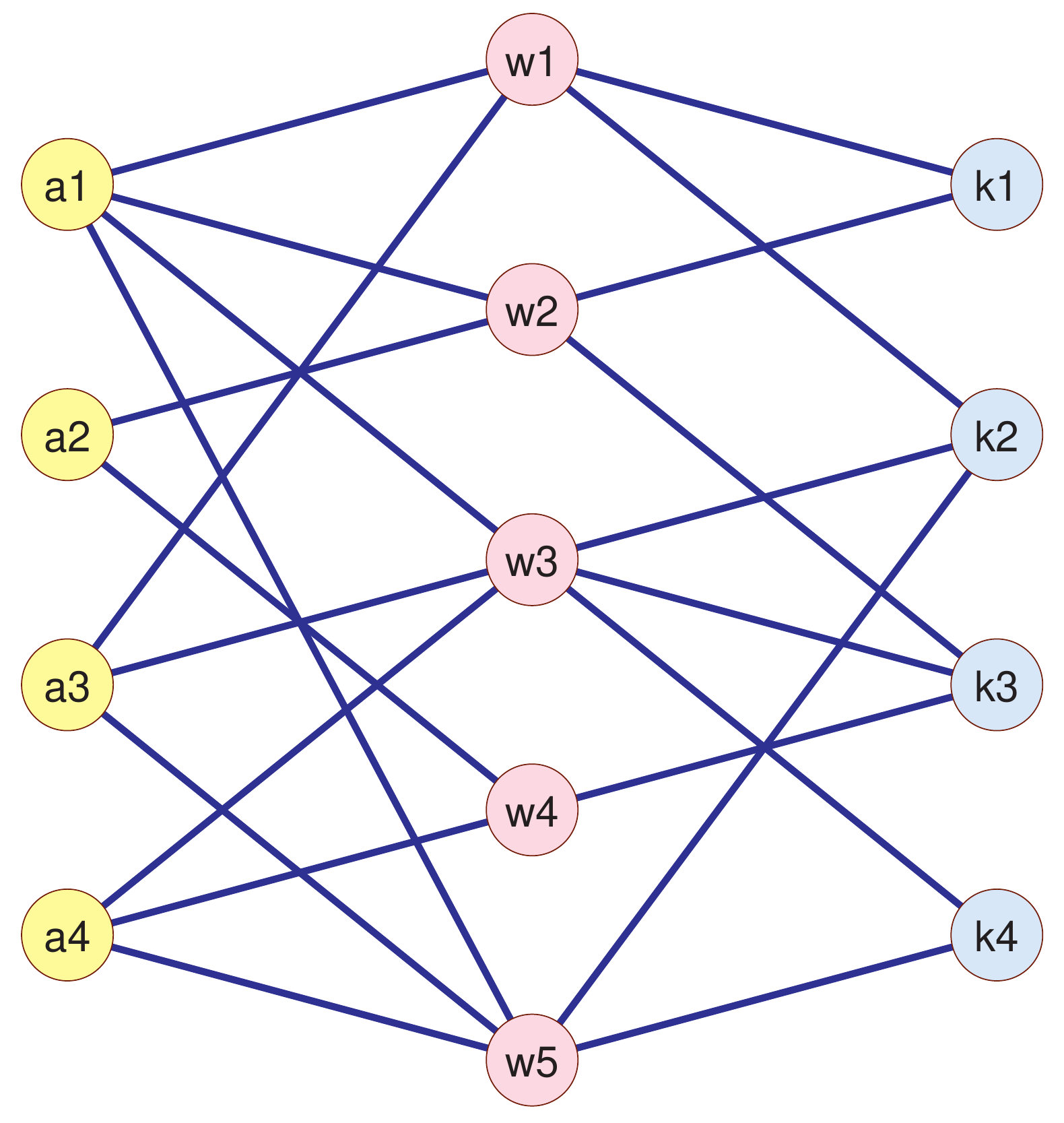}
\caption{$\WA^T \cdot \WK$}
\end{center}
\end{figure}

For vectors  $x = [x_1, x_2, \ldots, x_n]$ and $y = [y_1, y_2, \ldots, y_m]$ their \keyw{outer product} $x \circ y$ is defined as a matrix
$$x \circ y = [x_i \cdot y_j]_{n\times m}$$
then we can express the previous observation about the structure of product network as the \keyw{outer product decomposition} 
$$  \mathbf{C} = \mathbf{A} \cdot \mathbf{B} = \sum_k \Hm_k  \quad \mbox{where} \quad \Hm_k  =  \mathbf{A}[k,\cdot] \circ \mathbf{B}[k,\cdot]  $$
For binary (weights) networks we have $   \Hm_k  = K_{N^-_A(k),N_B(k)} $.

\paragraph{Example A:} As an example let us take the binary network matrices $\WA$ and $\WK$:

\[ 
\WA = \kbordermatrix{
        & a_1 & a_2 &  a_3 &  a_4  \\   
w_1  &  1   &   0    &   1    &    0  \\  
w_2  &  1   &   1    &   0    &    0   \\
w_3  &  1   &   0    &   1    &    1   \\ 
w_4  &  0   &   1    &   0    &    1   \\ 
w_5  &  1   &   0    &   1    &    1   },
\quad
\WK = \kbordermatrix{
        & k_1 & k_2 &  k_3 &  k_4  \\   
w_1 &   1   &  1    &   0   &    0    \\
w_2 &   1   &  0    &   1   &    0    \\
w_3 &   0   &  1    &   1   &    1    \\
w_4 &   0   &  0    &   1   &    0    \\
w_5 &   0   &  1    &   0   &    1    }
\]
and compute the product $\Hm = \WA^T \cdot \WK$. We get a network matrix $\Hm$ which can be decomposed as
\[
\kbordermatrix{
\Hm  & k_1 & k_2 & k_3 &  k_4  \\     
a_1 &  2   &   3   &   2   &   2    \\    	 
a_2 &  1   &   0   &   2   &   0    \\ 
a_3 &  1   &   3   &   1   &   2    \\     
a_4 &  0   &   2   &   2   &   2    }	   	 
= 
\kbordermatrix{
\Hm_1& k_1 & k_2 & k_3 &  k_4  \\ 
a_1  &  1   &   1   &   0   &   0    \\ 	 
a_2  &  0   &   0   &   0   &   0    \\ 
a_3  &  1   &   1   &   0   &   0    \\ 
a_4  &  0   &   0   &   0   &   0    } 
+
\kbordermatrix{
\Hm_2& k_1 & k_2 & k_3 &  k_4  \\ 
a_1 &   1   &   0   &  1   &    0    \\ 
a_2 &   1   &   0   &  1   &    0    \\  
a_3 &   0   &   0   &  0   &    0    \\ 
a_4 &   0   &   0   &  0   &    0    } 
+
\]
\[ 
\kbordermatrix{
\Hm_3& k_1 & k_2 & k_3 &  k_4  \\ 
a_1 &   0   &   1   &  1   &    1     \\
a_2 &   0   &   0   &  0   &    0     \\
a_3 &   0   &   1   &  1   &    1     \\
a_4 &   0   &   1   &  1   &    1    }
+
\kbordermatrix{
\Hm_4& k_1 & k_2 & k_3 &  k_4  \\ 
a_1 &   0   &   0   &   0  &    0    \\
a_2 &   0   &   0   &   1  &    0    \\
a_3 &   0   &   0   &   0  &    0    \\
a_4 &   0   &   0   &   1  &    0    }
+
\kbordermatrix{
\Hm_5& k_1 & k_2 & k_3 &  k_4  \\ 
a_1 &   0   &   1   &  0   &    1   \\
a_2 &   0   &   0   &  0   &    0   \\
a_3 &   0   &   1   &  0   &    1   \\
a_4 &   0   &   1   &  0   &    1   } \quad\strut
\]

\section{Derived networks}

We can use the multiplication to obtain new networks from existing \keyw{compatible} two-mode networks.
For example, from basic bibliographic networks $\WA$ and $\WK$ we get
\[ \AK = \WA^T \cdot \WK  \]
a network relating authors to keywords used in their works, and
\[\mathbf{Ca} = \WA^T \cdot \Ci \cdot \WA \]
 is a network of citations between authors.

Networks obtained from existing networks using some operations are called \keyw{derived} networks.
They are very important in analysis of collections of \keyw{linked} networks.

What is the meaning of the product network? In general we could consider weights, addition and multiplication over a selected semiring \citep{semi}. In this paper we will limit our attention to the traditional addition and multiplication of real numbers.

The weight $\AK[a,k]$ is equal to the number of times the author $a$ used the keyword $k$ in his/her works. 

The weight $\mathbf{Ca}[a,b]$ counts the number of times a work authored by the author $a$ is citing a work authored by the author $b$; or shorter, how many times the author $a$ cited the author $b$.

Using network multiplication we can also transform a given two-mode network, for example $\WA$, into corresponding ordinary one-mode networks (\keyw{projections})
\[ \mathbf{WW} = \WA \cdot \WA^T  \qquad \mbox{and} \qquad   \mathbf{AA} = \WA^T \cdot \WA \]
The obtained projections can be analyzed using standard network analysis methods. This is a traditional recipe how to analyze two-mode networks. Often the weights are not considered in the analysis; and when they are considered we have to be very careful about their meaning. 

The weight $\mathbf{WW}[p,q]$ is equal to the number of common authors of works $p$ and $q$. 

The weight $\mathbf{AA}[a,b]$ is equal to the number of works that author $a$ and $b$ coauthored. In a special case when $a=b$ it is equal to the number of works that the author $a$ wrote. The network $\mathbf{AA}$ is describing the \keyw{coauthorship}  (collaboration) between authors and is also denoted as $\mathbf{Co}$ -- the ``first'' coauthorship network.

In the paper \citet{fracBC} it was shown that there can be problems with the network $\mathbf{Co}$ when we try to use it for identifying the most collaborative authors. By the outer product decomposition the coauthorship network $\mathbf{Co}$  is composed of complete subgraphs on the set of work's coauthors. Works with many authors produce large complete subgraphs, thus bluring the collaboration structure, and are over-represented by its total weight. To see this, let $S_x = \sum_i x_i$ and  $S_y = \sum_j y_j$ then the \keyw{contribution} of the outer product $x\circ y$ is equal
$$T = \sum_{i,j} (x\circ y)_{ij} = \sum_i \sum_j x_i\cdot y_j = \sum_i  x_i\cdot \sum_j y_j = S_x \cdot S_y $$
In general each term $\Hm_w$ in the outer product decomposition of the product $\mathbf{C}$ has different total weight 
$T(\Hm_w) = \sum_{a,k} (\Hm_w)_{ak}$ leading to over-representation of works with large values. In the case of coautorship network $\mathbf{Co}$ we have $S(\WA[w,.]) = \outdeg_\WA(w)$ and therefore $T(\Hm_w) = \outdeg_\WA(w)^2$. To resolve the problem we apply the fractional approach.


\section{Fractional approach}

To make the contributions of all works equal we can apply the \keyw{fractional} approach by
normalizing the weights: setting $x' = x / S_x$ and $y' = y / S_y$ we get $S_{x'} = S_{y'} =1 $ and therefore
$T(\Hm'_w) = 1$ for all works $w$.

In the case of two-mode networks $\WA$ and $\WK$  we denote
\[  S^{\WA}_w =  \begin{cases}
   \sum_a \WA[w,a]       & \quad \outdeg_\WA(w) > 0\\
    1  & \quad \outdeg_\WA(w) = 0
  \end{cases}
\]
(and similarly $S^{\WK}_w$)
and define the \keyw{normalized} matrices
\[  \WA\mathbf{n} = \diag(\frac{1}{S^{\WA}_w}) \cdot \WA, \quad   \WK\mathbf{n} = \diag(\frac{1}{S^{\WK}_w}) \cdot \WK \]
In real life networks $\WA$ (or $\WK$) it can happen that some work has no author. In such a case $S^{\WA}_w = \sum_a \WA[w,a]  = 0$ which makes problems in the definition of the normalized network $\WA\mathbf{n}$. We can bypass the problem  by setting  $S^{\WA}_w = 1$, as we did in the above definition.

Then the \keyw{normalized product} matrix is
\[ \AK\mathbf{t} = \WA\mathbf{n}^T \cdot \WK\mathbf{n} \]
Denoting
$\displaystyle \mathbf{F}_w = \frac{1}{S^{\WA}_w S^{\WK}_w}  \Hm_w $  the outer product decomposition gets form
\[ \AK\mathbf{t} =  \sum_w \mathbf{F}_w  \]
Since
\[  T(\mathbf{F}_w) =   \begin{cases}
   1  & \quad (\outdeg_\WA(w) > 0)  \land  (\outdeg_\WK(w) > 0) \\
   0  & \quad \text{otherwise}
  \end{cases}
\]
we have further
\[  \sum_{a,k} \mathbf{F}[a,k] = \sum_{a,k} \sum_w \mathbf{F}_w[a,k] = \sum_w T(\mathbf{F}_w) = |W^+| \]
where $W^+ = \{ w \in W :  (\outdeg_\WA(w) > 0)  \land  (\outdeg_\WK(w) > 0) \}$.

In the network $\AK\mathbf{t}$ the contribution of each work to the bibliography is 1. These contributions are redistributed to arcs from authors to keywords.

\paragraph{Example B:}  For matrices from Example A we get the corresponding diagonal normalization matrices

\[
\diag(\frac{1}{S_w^{\WA}}) = \kbordermatrix{
       & w_1 & w_2 & w_3 &  w_4  & w_5  \\ 
w_1 & 1/2 &  0    &  0  &  0  &  0  \\    
w_2 &   0  & 1/2  &  0  &  0  &  0   \\   
w_3 &   0  &  0    & 1/3  &  0  &  0   \\  
w_4 &   0  &  0    &   0  & 1/2  &  0   \\  
w_5 &   0  &   0    &  0  &  0  &  1/3   }    
\]
\[
\diag(\frac{1}{S_w^{\WK}}) = \kbordermatrix{
       & w_1 & w_2 & w_3 &  w_4  & w_5  \\ 
w_1 & 1/2 &  0    &  0  &  0  &  0  \\    
w_2 &   0  & 1/2  &  0  &  0  &  0   \\   
w_3 &   0  &  0    & 1/3  &  0  &  0   \\  
w_4 &   0  &  0    &   0  &  1  &  0   \\  
w_5 &   0  &   0    &  0  &  0  &  1/2   }    
\]
compute the normalized matrices
\[
\WA\mathbf{n} = \kbordermatrix{
        & a_1 & a_2 &  a_3 &  a_4  \\  
w_1 & 1/2  &   0   &  1/2  &   0    \\    
w_2 & 1/2  & 1/2  &   0    &   0    \\  
w_3 & 1/3  &   0   & 1/3   & 1/3   \\  
w_4 &   0   & 1/2  &   0    & 1/2    \\    
w_5 & 1/3  &  0    & 1/3   & 1/3   } ,  
\quad
\WK\mathbf{n} = \kbordermatrix{
        & k_1 & k_2 &  k_3 &  k_4  \\   
w_1 &  1/2 & 1/2 &    0   &   0    \\
w_2 &  1/2 &   0  &  1/2  &   0    \\
w_3 &    0  & 1/3 &  1/3  & 1/3  \\
w_4 &    0  &   0  &    1    &  0    \\
w_5 &    0  & 1/2 &    0    & 1/2  },
\]
outer products such as
\[  \mathbf{F}_1 =
\kbordermatrix{
       & k_1 & k_2 & k_3 &  k_4  \\ 
a_1 & 1/4 & 1/4  &  0  &  0   \\    
a_2 &   0  &   0    &  0  &  0   \\   
a_3 & 1/4 & 1/4  &  0  &  0   \\  
a_4 &   0  &   0    &  0  &  0   }   
\qquad
\mathbf{F}_5 = 
\kbordermatrix{
       & k_1 & k_2 & k_3 &  k_4  \\ 
a_1 &  0   & 1/6  &   0   & 1/6  \\
a_2 &  0   &   0   &   0   &   0    \\
a_3 &  0   & 1/6  &   0   & 1/6   \\
a_4 &  0   & 1/6  &   0   & 1/6  }
\]
and finally the product matrix
\[ \AK\mathbf{t} =  \WA\mathbf{n}^T \cdot \WK\mathbf{n} =  \sum_{w=1}^5  \mathbf{F}_w =
\kbordermatrix{
       & k_1 & k_2 & k_3 &  k_4  \\         
a_1 & 0.50000 & 0.52778 & 0.36111 & 0.27778   \\   
a_2 & 0.25000 & 0.00000 & 0.75000 & 0.00000   \\      
a_3 & 0.25000 & 0.52778 & 0.11111 & 0.27778   \\  
a_4 & 0.00000 & 0.27778 & 0.61111 & 0.27778  }    
\]

\subsection{Linking through a network \label{through}}

Let a network $\mathbf{S}$  links works to works.  The derived network $\mathbf{WA}^T \cdot \mathbf{S} \cdot \mathbf{WA}$ links authors to authors \keyw{through} $\mathbf{S}$. Again, the normalization question has to be addressed. Among different options let us consider the derived networks defined as:
\[ \mathbf{C} = \mathbf{WAn}^T \cdot \mathbf{S} \cdot \mathbf{WAn} \]
It is easy to verify that:
\begin{itemize}
\item if $\mathbf{S}$ is symmetric, $\mathbf{S}^T = \mathbf{S}$, then also $\mathbf{C}$ is symmetric,   $\mathbf{C}^T = \mathbf{C}$;
\[ \mathbf{C}^T = ( \mathbf{WAn}^T \cdot \mathbf{S} \cdot \mathbf{WAn})^T =  \mathbf{WAn}^T \cdot \mathbf{S}^T \cdot (\mathbf{WAn}^T)^T = \mathbf{C} \]
\item if $W^+ =  \{ w \in W :  \outdeg_\WA(w) > 0 \} = W$, the total of weights of  $\mathbf{S}$ is redistributed in $\mathbf{C}$:
\[ T(\mathbf{C}) = \sum_{e \in L(\mathbf{C})} c(e) =  \sum_{e \in L(\mathbf{S})} s(e) = T(\mathbf{S})  \]
Since $\displaystyle \sum_{a \in A} wa[p,a] = \outdeg_\WA(p)$ and $\displaystyle wan[p,a] =
 \begin{cases}
    \frac{wa[p,a]}{\text{outdeg}_\WA(p)}  & \  \outdeg_\WA(p) > 0 \\
   0  & \  \text{otherwise}
  \end{cases}
$ \  we get
\[  T(\mathbf{C}) = \sum_{e \in L(\mathbf{C})} c(e) = \sum_{a \in A}\sum_{b \in A} c[a,b] = \sum_{a \in A}\sum_{b \in A} \sum_{p \in W}\sum_{q \in W} wan[p,a] \cdot s[p,q] \cdot wan[q,b] = \]
\[ = \sum_{p \in W^+}\sum_{q \in W^+} \frac{s[p,q]}{\outdeg_\WA(p)\outdeg_\WA(q)}  \sum_{a \in A} wa[p,a] \sum_{b \in A} wa[q,b] = \sum_{p \in W^+}\sum_{q \in W^+} s[p,q] \]
and finally, if $W^+ = W$
\[ \sum_{p \in W^+}\sum_{q \in W^+} s[p,q] = \sum_{e \in L(\mathbf{S})} s(e) = T(\mathbf{S})  \]
\end{itemize}

As special cases we get for normalized author's citation networks with $W^+ = W$: for $\mathbf{S} = \Ci$ 
\[  \sum_{a \in A}\sum_{b \in A} c[a,b] = \sum_{p \in W}\sum_{q \in W} ci[p,q] = | \Ci| \]
and for $\mathbf{S} = \mathbf{Cin}$
\[  \sum_{a \in A}\sum_{b \in A} c[a,b] = \sum_{p \in W}\sum_{q \in W:  \ \mathbf{outdeg_\Ci}(q) > 0} \frac{ci[p,q]}{\outdeg_\Ci(p)} = \sum_{q \in W:  \ \mathbf{outdeg_\Ci}(q) > 0} 1 = W_\Ci^+ \]

\subsection{Some notes\label{fa}}

\paragraph{A.} Instead of computing the normalized network $\WA\mathbf{n}$ from the network $\WA$ we could collect the data about the real proportion $wan[w,a]$ of the contribution of each author $a$ to a work $w$ such that $\WA\mathbf{n}$ is normalized: for every work $w$ it holds
\[  \sum_{a \in A} wan[w,a] \in \{0,1\} \]
Unfortunately in most cases such data are not available and we use the computed normalized weights as their estimates. Most of the results do not depend on the way the normalized network was obtained.

\paragraph{B.} In general a given network matrix $\WA$ can be normalized in two ways: \keyw{by rows}, as used in this section, and \keyw{by columns}
\[ \WA\mathbf{n'} = \WA \cdot \diag(\frac{1}{S^{\WA}_a}) \quad \text{where} \quad 
  S^{\WA}_a =  \begin{cases}
   \sum_w \WA[w,a]       & \  \indeg_\WA(a) > 0\\
    1  & \   \indeg_\WA(a) = 0
  \end{cases} \]
In the context of bibliographic networks its meaning does not make much sense.

\paragraph{C.} The network $\Co$ is symmetric: $ co_{ab} =  co_{ba} $. We need to compute only half of values
$ co_{ab}$, $a\le b$. The resulting network is undirected with weights $ co_{ab}$. 

\paragraph{D.} In the paper \citet{fracBC} the ``second'' coauthorship network  $\mathbf{Cn} = \WA^T\cdot \WA\mathbf{n}$ is considered. The weight $cn_{ab}$ is equal to the contribution of an author $a$ to works that (s)he wrote together with the author $b$. Using these weights the \keyw{selfsufficiency} of an author $a$ is defined as:
 $$\displaystyle S_a = \frac{cn_{aa}}{\indeg_{\WA}(a)}$$
and \keyw{collaborativness} of an author $a$ as its complementary measure $K_a = 1 - S_a$.

\paragraph{E.} In the ``third'' coauthorship network  $\mathbf{Cn} = \WA\mathbf{n}^T\cdot \WA\mathbf{n}$ the weight
$ct_{ab}$ is equal to the total fractional contribution of `collaboration' of authors $a$ and $b$ to works. Each work $w$ with 
 $S^\WA_w > 0$ contributes 1 to the total of weights in $\mathbf{Cn}$. This is the network to be used in analysis of collaboration between authors \citep{fracBC,fracLP,fracPM}.  To identify the most collaborative groups we can use methods such as $P_S$-cores and link islands \citep{Understand}.
 
The product  $\mathbf{Cn}$ is symmetric. Note \textbf{C} applies. We transform it to the corresponding undirected network -- pairs of opposite arcs are replaced by an edge with doubled weight. In analyses we usually analyze separately the vector of weights on loops (selfcontribution) and the network $\mathbf{Cn}$ without loops.
 
 
\paragraph{F.} An alternative normalization $\WA\mathbf{n'}$ of a binary autorship matrix $\WA$ was proposed in \citet{newman4}
 \[ wan'_{wa} =  \frac{wa_{wa} }{ \max(1,\outdeg_{\WA}(w)-1)} \]
in which only collaboration with coauthors is considered -- no selfcollaboration. Note that using the network construction proposed on page 5 of \cite{newman1} we get a network in which works with many coauthors are still over-represented. The same idea is used in the fractional counting co-authorship matrix $\mathbf{U}^*$ proposed in equation (5) in \citet{fracPWE}. 

To treat all works equally using the Newman's normalization the ``fourth'' coauthorship network was proposed in \citet{fracCB}. To compute it we first compute
\[ \mathbf{Ct'} = \WA\mathbf{n}^T \cdot \WA\mathbf{n'} \]
The weight $ct'_{ab}$ is equal to the total contribution of ``strict collaboration'' of authors $a$ and $b$ to works. The obtained product is symmetric. Again note \textbf{C} applies. We transform it to the corresponding undirected network -- pairs of opposite arcs are replaced by an edge with doubled weight. The loops are removed. The contribution of each work with at least two coauthors is equal to 1. A kind of the outer product decomposition exists also for the network $\mathbf{Ct'}$ with a diagonal set to 0.


\section{Bibliographic Coupling and Co-citation}

\begin{figure}
\begin{center}
\includegraphics[width=65mm]{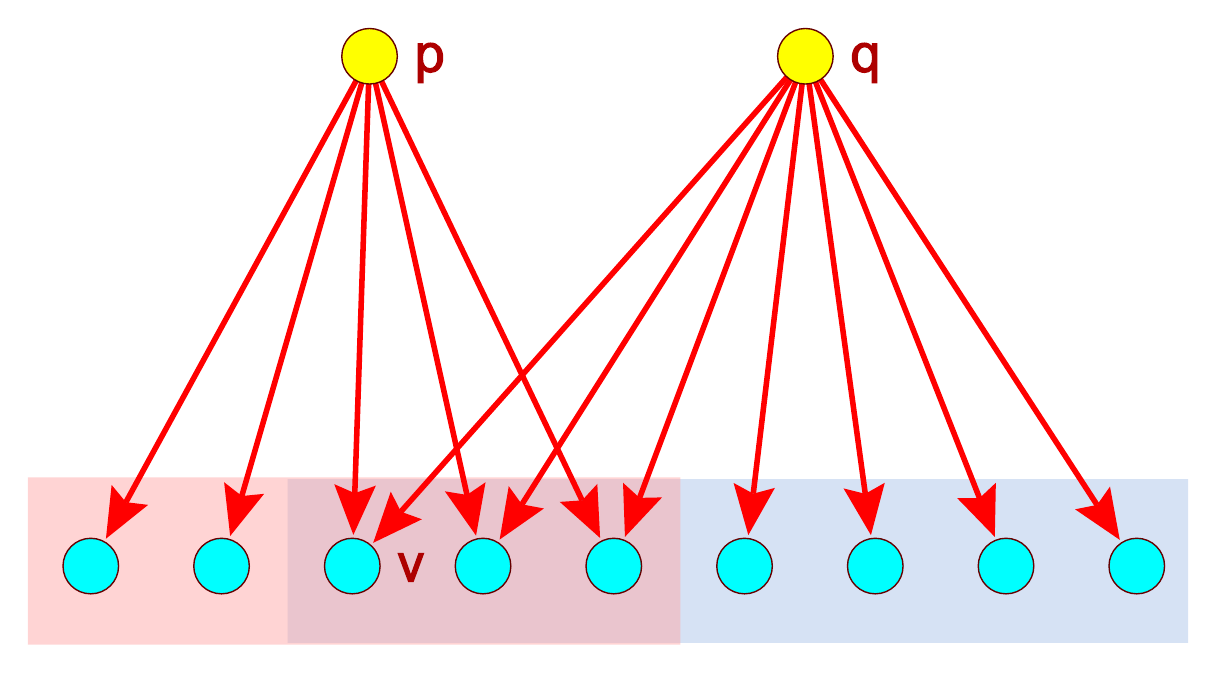} \qquad \includegraphics[width=65mm]{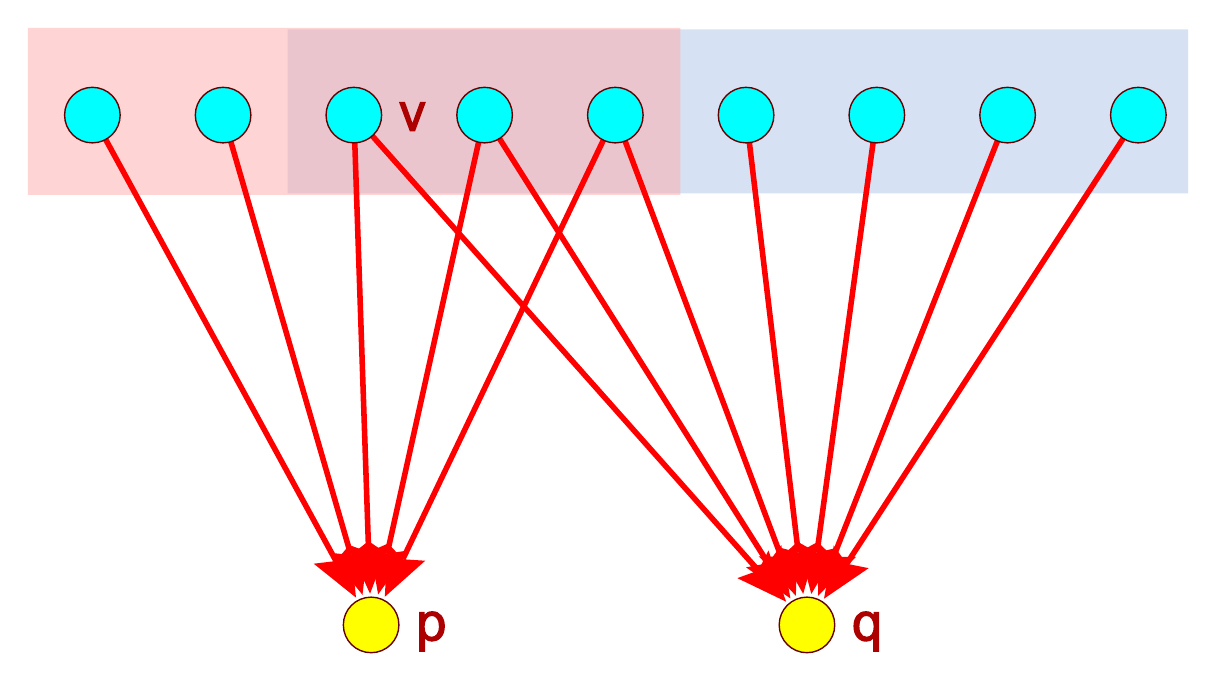}
\end{center}
\caption{Bibliographic coupling (left) and Co-citation (right)}\label{cico}
\end{figure}

Bibliographic coupling occurs when two works each cite a third work in their bibliographies, see Figure~\ref{cico}, left. The idea was introduced by Kessler (1963) and has been used extensively since then. See figure where two citing works, $p$ and $q$, are shown. Work $p$ cites five works and $q$ cites seven works. The key idea is that there are three works cited by both $p$ and $q$. This suggests some content communality for the three works cited by both $p$ and $q$. Having more works citing pairs of prior works increases the likelihood of them sharing content.

We assume that the citation relation means 
$ p\ \mathbf{Ci}\ q \equiv \mbox{~work~} p \mbox{~cites work~} q $.
Then the \keyw{bibliographic coupling} network $\mathbf{biCo}$ can be determined as
\[ \mathbf{biCo} = \mathbf{Ci} * \mathbf{Ci}^T \]
The weight $bico_{pq} $ is equal to the number of works cited by both works $p$ and $q$; $bico_{pq}= |  \mathbf{Ci}(p) \cap  \mathbf{Ci}(q) |$. 
Bibliographic coupling weights are symmetric:  $bico_{pq} = bico_{qp}$:
\[ \mathbf{biCo}^T = (\mathbf{Ci} \cdot \mathbf{Ci}^T)^T =  \mathbf{Ci} \cdot \mathbf{Ci}^T =   \mathbf{biCo} \]


Co-citation is a concept with strong parallels with bibliographic coupling (Small and Marshakova 1973), see Figure~\ref{cico}, right. The focus is on the extent to which works are co-cited by later works. The basic intuition is that the more earlier works are cited, the higher the likelihood that they have common content. 
The \keyw{co-citation}   network $\mathbf{coCi}$ can be determined as
\[ \mathbf{coCi} = \mathbf{Ci}^T \cdot \mathbf{Ci} . \]
The weight $coci_{pq}$ is equal to the number of works citing both works $p$ and $q$. The network $\mathbf{coCi}$ is symmetric $coci_{pq} = coci_{qp}$:
\[ \mathbf{coCi}^T = (\mathbf{Ci}^T \cdot \mathbf{Ci})^T = \mathbf{Ci}^T \cdot \mathbf{Ci} =  \mathbf{coCi} \]
An important property of co-citation is that $\mathbf{coCi}(\Ci) = \mathbf{biCo}(\Ci^T) $:
\[  \mathbf{biCo}(\Ci^T) = \Ci^T \cdot (\Ci^T)^T = \Ci^T \cdot \Ci = \mathbf{coCi}(\Ci) \]
Therefore the constructions proposed for bibliographic coupling can be applied also for co-citation.

What about normalizations? Searching for the most coupled works we have again  problems with works with many citations, especially with review papers. To neutralize their impact we can introduce normalized measures.
The fractional approach works fine for normalized co-citation
\[ \mathbf{CoCit} = \mathbf{Cin}^T \cdot \mathbf{Cin} \]
where $\Ci\mathbf{n} = \mathbf{D} \cdot \mathbf{Ci}$ and $\mathbf{D} = \diag(\frac{1}{\max(1,\text{outdeg}(p))})$.  $\mathbf{D}^T = \mathbf{D}$ . In the normalized network every work has value 1 and it is equally distributed to all cited works.

The fractional approach can not bi directly applied to bibliographic coupling -- to get the outer product decomposition work we would need to normalize $\Ci$ by columns -- a cited work has value 1 which is distributed equally to the citing works -- the most cited works give the least.  This is against our intuition. To construct a reasonable measure we can proceed as follows.  Let us first look at
\[ \mathbf{biC} =  \Ci\mathbf{n} \cdot \mathbf{Ci}^T \]
we have
\[  \mathbf{biC} = (\mathbf{D} \cdot \mathbf{Ci}) \cdot \mathbf{Ci}^T = \mathbf{D} \cdot \mathbf{biCo} \]
\[ \mathbf{biC}^T = (\mathbf{D} \cdot \mathbf{biCo})^T =  \mathbf{biCo}^T \cdot \mathbf{D}^T =  \mathbf{biCo} \cdot \mathbf{D} \]
For $\mathbf{Ci}(p) \ne \emptyset$ and  $\mathbf{Ci}(q) \ne \emptyset$  it holds 
\[  \mathbf{biC}_{pq} = \frac{|\mathbf{Ci}(p) \cap \mathbf{Ci}(q)|}{|\mathbf{Ci}(p)|} \quad \mbox{and} \quad
 \mathbf{biC}_{qp} = \frac{|\mathbf{Ci}(p) \cap \mathbf{Ci}(q)|}{|\mathbf{Ci}(q)|} =  \mathbf{biC}_{pq}^T \]
and $\mathbf{biC}_{pq} \in [0,1]$. $ \mathbf{biC}_{pq}$ is the proportion of its references that the work $p$ shares with the work $q$. The network $\mathbf{biC}$ is not symmetric. We have different options to  construct  normalized symmetric measures such as
\[ \mathbf{biCoa}_{pq} = \frac{1}{2}( \mathbf{biC}_{pq} +  \mathbf{biC}_{qp} )   \quad \mbox{Average}\]
\[ \mathbf{biCom}_{pq} = \min( \mathbf{biC}_{pq}, \mathbf{biC}_{qp} )  \quad \mbox{Minimum} \]
\[ \mathbf{biCoM}_{pq} = \max( \mathbf{biC}_{pq}, \mathbf{biC}_{qp} )  \quad \mbox{Maximum} \]
or, may be more interesting
\[ \mathbf{biCog}_{pq} = \sqrt{ \mathbf{biC}_{pq}\cdot \mathbf{biC}_{qp}} =
\frac{|\mathbf{Ci}(p) \cap  \mathbf{Ci}(q)|}{\sqrt{ |\mathbf{Ci}(p)| \cdot |\mathbf{Ci}(q) |} } \quad \begin{array}{l}\mbox{Geometric mean}\\\mbox{Salton cosinus}\end{array} \] 
\[ \mathbf{biCoh}_{pq} = 2\cdot ( \mathbf{biC}_{pq}^{-1} +  \mathbf{biC}_{qp}^{-1} )^{-1} =
\frac{ 2 |\mathbf{Ci}(p) \cap  \mathbf{Ci}(q)|}{ |  \mathbf{Ci}(p)| +  |\mathbf{Ci}(q) |}  \quad \mbox{Harmonic mean} \]
\[ \mathbf{biCoj}_{pq} = ( \mathbf{biC}_{pq}^{-1} +  \mathbf{biC}_{qp}^{-1} - 1)^{-1} =
\frac{ |\mathbf{Ci}(p) \cap  \mathbf{Ci}(q)|}{ |  \mathbf{Ci}(p) \cup \mathbf{Ci}(q) |} \quad \mbox{Jaccard index} \]
All these measures are similarities.

It is easy to verify that $biCoX_{pq} \in [0,1]$ and: $biCoX_{pq} = 1$ iff the works $p$ and $q$ are
referencing the same works, $\mathbf{Ci}(p) = \mathbf{Ci}(q)$.\medskip

From $m \leq H \leq G \leq A \leq M$ and  $J \leq m$, ($ \frac{|P \cap Q|}{|P \cup Q|} \leq \min(\frac{|P \cap Q|}{|P|} ,\frac{|P \cap Q|}{|Q|} )$) we get
\[  \mathbf{biCoj}_{pq} \leq  \mathbf{biCom}_{pq} \leq \mathbf{biCoh}_{pq} \leq \mathbf{biCog}_{pq} \leq  \mathbf{biCoa}_{pq}  \leq  \mathbf{biCoM}_{pq}\]
The equalities hold iff $\mathbf{Ci}(p) = \mathbf{Ci}(q)$. 

To get a dissimilarity we can use transformations $dis = 1 - sim$ or $dis = \frac{1}{sim} - 1$ or $dis = - \log sim$. For example
\[ \mathbf{biCod}_{pq} = 1 -  \mathbf{biCoj}_{pq} =
\frac{ |\mathbf{Ci}(p) \oplus \mathbf{Ci}(q)|}{ |  \mathbf{Ci}(p) \cup \mathbf{Ci}(q) |}  \quad \mbox{Jaccard distance} \]
where $\oplus$ denotes the symmetric difference of sets.

Bibliographic coupling and co-citation networks are linking works to works. To get linking between authors, journals or keywords considering citation similarity we can apply the construction from Subsection~\ref{through} to the normalized co-citation or bibliographic coupling network.

\section{Conclusions}

In the paper we presented an attempt to provide a foundation of fractional approach to biblimetric networks based on the outer product decomposition of product networks. We also discussed the fractional approach to bibliographic coupling and co-citation networks. The results of application of the proposed methods to real bibliographic data will be presented in separate papers.

All described computations can be done efficiently  in program Pajek \citep{pajek} using macros such us:  
\texttt{norm1} -- normalized 1-mode network, 
\texttt{norm2} -- normalized 2-mode network,
\texttt{norm2p} -- Newman's normalization of a 2-mode network, 
\texttt{biCo} -- bibliographic coupling network, and
\texttt{biCon} -- normalized bibliographic coupling network, available at GitHub \citep{biblio}.

\section*{Acknowledgments}

The paper is based on presentations on 
1274. Sredin seminar, IMFM, Ljubljana, 29. March 2017;
NetGloW 2018,  St Petersburg, July 4-6, 2018; and
COMPSTAT 2018, Iasi, Romania, August 28-31, 2018.
 
This work is supported in part by the Slovenian Research Agency (research program P1-0294 and research projects J1-9187, J7-8279 and BI-US/17-18-045),  project CRoNoS (COST Action IC1408) and by Russian Academic Excellence Project '5-100'.


\end{document}